# Constraining Nonminimal $f(T)$ Gravity from Primordial Nucleosynthesis to Late-Universe Observations

Yahia Al-Omar [1], Majida Nahili [1,2], and Nidal Chamoun[3,4]*

[1] Department of Physics, Faculty of Sciences, Damascus University, Damascus, Syria
[2] Faculty of Pharmacy, Syrian Private University, Damascus, Syria
[3] Department of Statistics, Faculty of Sciences, Damascus University, Damascus, Syria
[4] CASP, Antioch Syrian University, Maaret Saidnaya, Damascus, Syria

* Corresponding author
E-mail: yahia.alomar@damascusuniversity.edu.sy, majeda.nahili@damascusuniversity.ed.sy, and chamoun@uni-bonn.de

## Abstract

We present a multi-epoch test of $f(T)$ gravity with nonminimal torsion–matter coupling, combining early- and late-Universe observations. At the MeV scale, Big-Bang Nucleosynthesis constrains the fractional variation of the weak freeze-out temperature, $|\delta \mathcal{T}_f/\mathcal{T}_f|$, thereby mapping light-element abundances into limits on deviations from the standard expansion history. At low redshift, we confront the model with type Ia supernovae, baryon acoustic oscillations, and cosmic-chronometer data, which respectively probe distances, the late-time standard ruler, and the Hubble rate. Independent analyses highlight the complementary roles of each dataset, while a joint SNe Ia + BAO + CC fit breaks degeneracies and yields the tightest combined bounds. As an illustration, we examine two representative torsion-modified gravity scenarios: BBN strongly limits large departures from standard cosmology, whereas late-time probes remain compatible with a near-ΛCDM background. This unified approach demonstrates the power of linking early-Universe nuclear physics with precision cosmological observables in assessing torsional extensions of gravity.

## 1. Introduction

For more than two decades, the ΛCDM model has served as the standard framework of modern cosmology, successfully explaining the Universe's accelerated expansion, structure formation, and CMB anisotropies [1]. It employs dark energy and cold dark matter to fit a wide range of observations with remarkable precision. However, the model's reliance on poorly understood dark energy and dark matter, combined with emerging cosmological tensions, raises questions about its completeness [2].

The most significant issue is the Hubble tension [3], in which early-Universe measurements inferred from the CMB disagree with late-Universe [4], local distance-ladder determinations by more than $4\sigma$ [5]. Another important discrepancy, the $S_8$ tension, reflects a mismatch in the amplitude of matter fluctuations between early-Universe CMB predictions and late-time gravitational-lensing surveys. These conflicts suggest either shortcomings within ΛCDM or unresolved observational systematics. Such tensions motivate the exploration of new physics beyond the standard cosmological model [6]. Rather than introducing additional exotic components such as dark energy, modifying gravity on large scales offers a more unified framework that can simultaneously address both dark energy and dark matter phenomenology [7].

General Relativity (GR) describes gravity through spacetime curvature, whereas teleparallel gravity provides an alternative formulation in which gravity arises from torsion, using a curvature-free connection. Originally explored by Einstein, this framework leads to the Teleparallel Equivalent of General Relativity (TEGR), which reproduces GR's field equations but differs in its geometric structure and extensions [8]. By generalizing the torsion scalar $T$ into an arbitrary function $f(T)$, one obtains $f(T)$ gravity—a modified-gravity theory characterized by simpler, second-order field equations that avoid certain instabilities present in other extensions such as $f(R)$ gravity [9].

Nonminimal coupling (NMC) in $f(T)$ gravity links the torsion scalar to matter fields, addressing issues such as the strong-coupling problem, in which a hidden scalar degree of freedom can appear in perturbations. This coupling is not a superficial modification but a theoretically motivated mechanism that resolves deeper structural challenges [10]. By introducing a scalar–matter interaction, NMC models clarify the behavior of the hidden degree of freedom, unify geometric and matter sectors, reopen theories previously limited by gravitational-wave constraints while preserving the observed wave speed, and can account for cosmic acceleration without a cosmological constant—all while remaining compatible with current data [11].

A diverse set of observational probes plays a crucial role in this task. Big-Bang Nucleosynthesis (BBN) refers to the nuclear reactions in the first few minutes after the Big Bang that formed light elements such as hydrogen, helium-4, and deuterium [12]. The process is highly sensitive to the Universe's expansion rate, meaning that any deviation from standard cosmology could alter the primordial element abundances [13]. Recent studies have tightened constraints on the baryon abundance using improved observational data and more precise nuclear reaction rates, particularly for deuterium [14]. Beyond testing the standard cosmological model, BBN also provides a powerful probe of new physics: numerous studies show that modified gravity scenarios—including scalar–tensor theories, higher-dimensional frameworks, and $f(T)$ gravity—are strongly constrained by the requirement of consistency with light-element abundances [15–19].

Moving to intermediate and low redshifts, Type Ia supernovae (SNe Ia) act as "standardizable candles" and provide measurements of luminosity distances across cosmic history [20–22] They are essential in probing cosmic acceleration and were historically central to the discovery of dark energy. Baryon Acoustic Oscillations (BAO) complement this picture by serving as a standard ruler, constraining the expansion rate and angular diameter distance with high precision [23,24]. Furthermore, Cosmic chronometers (CC) act as differential age indicators, providing direct, model-independent measurements of the Hubble parameter up to intermediate redshifts [25,26]. These late-time observables have been widely used to test and constrain modified gravity scenarios. Several studies combining SNe Ia, BAO, and CC data have constrained the free parameters of $f(T)$ gravity allowing only mild deviations from ΛCDM while remaining observationally viable [27–29]. Extented torsional models such as $f(T, T_G)$ have similarly been shown to fit supernovae, BAO, and $H(z)$ data, supporting their ability to drive late-time acceleration without a cosmological constant [30]. The latest studies also combine DESI BAO and DES-SN supernovae data, showing that $f(T)$ models can produce $H_0$ values closer to local determinations while remaining compatible with late-time observations [31]. Our joint-likelihood analysis using SNe Ia, BAO, and CC provides a timely test of a torsion–matter coupling mechanism within modified teleparallel gravity, directly comparable with recent observational constraints.

Each probe, constraining different epochs and aspects of cosmic evolution, complements the others: BBN tests the early Universe, SNe Ia and BAO probe late-time acceleration, and CC directly trace $H(z)$ across cosmic time. By analyzing these datasets individually and in combination, one can test the viability of modified teleparallel gravity against ΛCDM in a systematic way.

This study progresses from early-Universe foundations to observational tests of torsion-based gravity. Section 2 explores the teleparallel formulation of gravity. Section 3 reviews BBN and its observational constraints. Section 4 presents the late-time observational strategy and statistical methods for parameter estimation. Section 5 integrates early- and late-time constraints to test $f(T)$ scenarios against ΛCDM and contemporary data. Section 6 summarizes the key findings and their implications.

## 2. Teleparallel Formalism with Nonminimal Torsion–Matter Coupling

Teleparallel gravity reinterprets General Relativity by replacing curvature with torsion as the origin of gravitational interactions [32]. In this framework, tetrad fields $e_\mu^A$ are fundamental, relating the spacetime metric to the flat Minkowski metric via $g_{\mu\nu} = \eta_{AB} e_\mu^A e_\nu^B$. The Weitzenböck connection, $\Gamma^\lambda{}_{\mu\nu} = e_A{}^\lambda \partial_\nu e^A{}_\mu$, has vanishing curvature but nonzero torsion, giving the torsion tensor $T^\lambda{}_{\mu\nu} = \Gamma^\lambda{}_{\nu\mu} - \Gamma^\lambda{}_{\mu\nu}$. The superpotential, $S_\rho{}^{\mu\nu} = \frac{1}{2}(K^{\mu\nu}{}_\rho + \delta_\rho^\mu T^{\alpha\nu}{}_\alpha - \delta_\rho^\nu T^{\alpha\mu}{}_\alpha)$, constructed from the contorsion tensor $K^\lambda{}_{\mu\nu} = \Gamma^\lambda{}_{\mu\nu} - \tilde{\Gamma}^\lambda{}_{\mu\nu}$, where $\tilde{\Gamma}^\lambda{}_{\mu\nu}$ is the Levi Civita connection, combines with the torsion tensor to form the torsion scalar, $T = S_\rho{}^{\mu\nu} T^\rho{}_{\mu\nu}$, analogous to the Ricci scalar in curvature-based gravity [33].

The simplest action in this formulation is TEGR given by,

$$S_{\text{TEGR}} = \frac{1}{2\kappa^2} \int d^4x\, e [T + \mathcal{L}_m], \tag{1}$$

where $\kappa^2 = 8\pi G$, $e = \det(e^A_\mu) = \sqrt{-g}$, and $\mathcal{L}_m$ is the matter Lagrangian density. Varying this action with respect to the tetrad reproduces the Einstein field equations, demonstrating the equivalence between TEGR and GR [34].

A standard and well-motivated extension of TEGR is to promote the torsion scalar $T$ to a general function $f(T)$, leading to the action

$$S_{f(T)} = \frac{1}{2\kappa^2} \int d^4x\, e [T + f(T) + \mathcal{L}_m]. \tag{2}$$

This promotion of $T$ to $f(T)$ introduces new dynamical degrees of freedom beyond those of General Relativity, establishing a torsional counterpart to $f(R)$ gravity. It has been extensively explored in cosmology, particularly as a candidate to drive late-time cosmic acceleration without a cosmological constant [35]. Variation of the action (2) with respect to the tetrad field yields the corresponding field equations

$$(1 + f')[e^{-1} \partial_\mu(e e^\rho_A S_\rho{}^{\nu\mu}) - e^\lambda_A T^\rho{}_{\mu\lambda} S_\rho{}^{\mu\nu}] + e^\rho_A S_\rho{}^{\nu\mu} \partial_\mu T f'' + \frac{1}{4} e^\nu_A [T + f] = \kappa^2/2\; e^\lambda_A \Theta^\nu_\lambda, \tag{3}$$

where primes denote derivatives with respect to $T$, and $\Theta^\nu_\lambda$ is the energy–momentum tensor.

A further generalization introduces a direct interaction between torsion and matter through a nonminimal coupling (NMC) [36]. The action takes the form

$$S = \frac{1}{2\kappa^2} \int d^4x\, e\, [T + f_1(T) + (1 + \lambda f_2(T))\mathcal{L}_m], \tag{4}$$

where $f_1(T)$ and $f_2(T)$ are arbitrary functions of $T$, and $\lambda$ is a coupling constant controlling the strength of the torsion–matter interaction. In the limit $\lambda \to 0$, the theory reduces to the familiar $f(T)$ scenario, while for $f_1(T) = T$, and $f_2(T) = 0$, one recovers TEGR. Varying (4) yields the modified field equations:

$$(1 + f_1' + \lambda f_2' \mathcal{L}_m)[e^{-1} \partial_\mu(e e^\alpha_A S_\alpha{}^{\rho\mu}) - e^\alpha_A T^\mu{}_{\nu\alpha} S_\mu{}^{\nu\rho}] + (f_1'' + \lambda f_2'' \mathcal{L}_m) \partial_\mu T e^\alpha_A S_\alpha{}^{\rho\mu} + \frac{1}{4} e^\rho_A (f_1 + T) - \frac{1}{4} \lambda f_2' \partial_\mu T e^\alpha_A \frac{\partial \mathcal{L}_m}{\partial \partial_\mu e^A_\rho} + \lambda f_2' e^\alpha_A S_\alpha{}^{\rho\mu} \partial_\mu \mathcal{L}_m = \frac{1}{2}\kappa^2 (1 + \lambda f_2) \Theta^\rho_A. \tag{5}$$

For a spatially flat Friedmann–Lemaître–Robertson–Walker (FLRW) universe with tetrad $e^A_\mu = \text{diag}(1, a(t), a(t), a(t))$, the torsion scalar reduces to $T = -6H^2$, where $H = \dot{a}/a$ is the Hubble parameter. In this background, the modified Friedmann equations take the form [37]:

$$H^2 = \frac{\kappa^2}{3}[1 + \lambda(f_2 + 12H^2 f_2')]\rho_m - \frac{1}{6}(f_1 + 12H^2 f_1'), \tag{6}$$

$$\dot{H} = -\frac{\kappa^2 (\rho_m + p_m)/2\, [1 + \lambda(f_2 + 12H^2 f_2')]}{1 + f_1' - 12H^2 f_1'' - 16\pi G \lambda \rho_m (f_2' - 12H^2 f_2'')}, \tag{7}$$

with $\rho_m$ and $p_m$ being the matter density and pressure, respectively. The effective interaction between torsion and matter can be interpreted as an additional dark-energy-like contribution to the cosmic dynamics, with corresponding energy-density and pressure:

$$\rho_{DE} = -\frac{1}{2\kappa^2}(f_1 + 12H^2 f_1') + \lambda \rho_m (f_2 + 12H^2 f_2'), \tag{8}$$

$$p_{DE} = (\rho_m + p_m) \times \left[ \frac{1 + \lambda(f_2 + 12H^2 f_2')}{1 + f_1' - 12H^2 f_1'' - 16\pi G \lambda \rho_m (f_2' - 12H^2 f_2'')} - 1 \right] + \frac{1}{2\kappa^2}(f_1 + 12H^2 f_1') \quad (9)$$
$$- \lambda \rho_m (f_2 + 12H^2 f_2').$$

This construction alters both the gravitational field equations and the conservation law of the matter sector: the matter energy–momentum tensor is no longer independently conserved, and the resulting torsion–matter coupling generates an additional force acting on matter fields. These deviations provide a natural and testable avenue for discriminating this class of models from the standard ΛCDM cosmology as well as from other modified-gravity frameworks.

## 3. Big Bang Nucleosynthesis and Parameter Bounds

BBN provides one of the earliest and most sensitive tests of cosmological models, linking high-energy physics to observed primordial element abundances. Occurring within the first few minutes after the Big Bang at temperatures of $0.1 - 1$ MeV, it leads to the formation of light nuclei such as $^4$He, deuterium, and trace amounts of $^3$He and $^7$Li [38,39]. Because the process is governed by the competition between weak interaction rates and the Hubble expansion rate, BBN serves as a powerful probe of non-standard scenarios, including modified gravity.

In the standard cosmological model, the Universe is radiation dominated during BBN, with radiation energy density

$$\rho_r = \frac{\pi^2}{30} g_* \mathcal{T}^4, \quad (10)$$

where $\mathcal{T}$ is the temperature and $g_*$ is the effective number of relativistic degrees of freedom. At $\mathcal{T} \sim 1$ MeV, one finds $g_* \simeq 10.75$, accounting for photons, electron–positron pairs, and three neutrino species [40]. The corresponding Hubble rate in GR is

$$H_{GR}^2 \approx \frac{1}{3M_P^2} \rho_r, \quad (11)$$

with $M_P = (8\pi G)^{-1/2}$ the reduced Planck mass. At high temperatures, neutrons and protons remain in chemical equilibrium through weak processes such as

$$n + \nu_e \leftrightarrow p + e^-, \quad n + e^+ \leftrightarrow p + \bar{\nu}_e, \quad n \rightarrow p + e^- + \bar{\nu}_e. \quad (12)$$

The rate of these reactions can be approximated by

$$\lambda_{tot}(\mathcal{T}) \approx c_q \mathcal{T}_f^5, \quad c_q = 9.8 \times 10^{-10} \, GeV^{-4}, \quad (13)$$

reflecting the temperature dependence of weak processes [41]. Freeze-out occurs when $H \sim \lambda_{tot}$, which in standard cosmology happens around $\mathcal{T}_f \sim 0.8$ MeV. At freeze-out, the neutron-to-proton ratio is approximately

$$\frac{n_n}{n_p} \simeq e^{-\frac{Q}{\mathcal{T}_f}}, \quad Q = m_n - m_p = 1.293 \text{ MeV}, \quad (14)$$

and subsequently decreases due to neutron decay, with lifetime $\tau_n \simeq 877$ s [42]. The most precisely measured product of BBN is the helium-4 mass fraction $Y_p$, which depends sensitively on the neutron–proton ratio at freeze-out. A useful approximation is

$$Y_p \simeq \frac{2x_f}{1 + x_f} \exp\left[-\frac{t_n - t_f}{\tau_n}\right], \quad (15)$$

where $x_f = n_n/n_p$ at freeze-out, $t_f$ is the freeze-out time, and $t_n$ marks the onset of nucleosynthesis when deuterium becomes stable. Current observations give $Y_p = 0.2449 \pm 0.0040$ [43], providing a tight constraint on deviations from the standard expansion rate.

In theories beyond GR, the expansion rate is modified. A common parametrization is

$$H^2 \approx \frac{1}{3M_P^2} \rho_r \left(1 + \frac{\rho_{DE}}{\rho_r}\right), \tag{16}$$

where $\rho_{DE}$ represents additional contributions from new physics (e.g., modified gravity). During BBN, $\rho_{DE} \ll \rho_r$, so the deviation from the GR expansion rate is

$$\Delta H \approx \frac{\rho_{DE}}{\rho_r} \frac{H_{GR}}{2}, \tag{17}$$

which shifts the freeze-out temperature according to

$$\frac{\delta \mathcal{T}_f}{\mathcal{T}_f} = \frac{\rho_{DE}}{\rho_r} \frac{H_{GR}}{10 c_q \mathcal{T}_f^5}. \tag{18}$$

This perturbation affects the neutron–proton ratio and therefore the helium abundance. To remain consistent with observations, the shift must satisfy

$$\left|\frac{\delta \mathcal{T}_f}{\mathcal{T}_f}\right| < 4.7 \times 10^{-4}. \tag{19}$$

This constraint translates directly into bounds on modifications of gravity at MeV energies. BBN therefore serves as a "cosmic laboratory," probing gravitational physics during an epoch inaccessible to terrestrial experiments. Modern analyses combine BBN, CMB measurements, and primordial abundance data, making it one of the most stringent early-universe tests of extended gravity scenarios [44,45].

## 4. Observational Constraints from CC, SNe Ia, and BAO

To place robust constraints on the cosmic expansion history and on the dynamics of modified gravity models, we employ three complementary and largely independent observational probes: CC, SNe Ia and BAO. Each of these datasets provides access to different aspects of the expansion history, and their joint analysis significantly improves the precision of cosmological parameter estimation.

The CC method provides a direct, model-independent determination of the Hubble parameter $H(z)$ through differential age measurements of passively evolving early-type galaxies that formed at high redshift ($z \gtrsim 2$) [46,47]. By comparing the relative ages of such galaxies at nearby redshifts, one obtains

$$H(z) = -\frac{1}{1+z} \frac{\Delta z}{\Delta t}, \tag{20}$$

where $\Delta t$ is the age difference between galaxies separated by a small redshift interval $\Delta z$.

In this work, we employ a compilation of 46 CC measurements spanning $0.0 \leq z \leq 2.36$ [48], thereby covering epochs from the matter-dominated era to the present accelerated phase. In particular, the present-day Hubble constant $H_0$ remains a central parameter of interest, especially in light of the long-standing tension between early- and late-Universe measurements: Planck CMB observations favor $H_0 = 67.4 \pm 0.5$ km s$^{-1}$Mpc$^{-1}$[4], whereas local distance-ladder calibrations suggest a higher value, $H_0 = 73.04 \pm 1.04$ km s$^{-1}$Mpc$^{-1}$ [5]. The associated likelihood is computed via

$$\chi^2_{CC} = \sum_{i=1}^{46} \frac{\left[H_i^{th}(\theta_s, z_i) - H_i^{obs}(z_i)\right]^2}{\sigma^2_{CC}(z_i)}, \tag{21}$$

where $H^{\text{obs}}(z_i)$ and $\sigma_{\text{CC}}(z_i)$ denote the observed values and their uncertainties, and $H^{\text{th}}(\theta, z_i)$ represents the theoretical model prediction, and $\theta$ is the set of free cosmological parameters.

Type Ia supernovae act as standardizable candles due to their relatively uniform intrinsic luminosity near the Chandrasekhar mass limit [49–51]. The observed apparent magnitude $m$ is related to the absolute magnitude $M$ through the distance modulus

$$\mu = m - M = 5 \, log_{10}\left(\frac{d_L}{Mpc}\right) + 25, \tag{22}$$

where the luminosity distance is

$$d_L(z) = (1+z) \int_0^z dz' \frac{c}{H(z')}. \tag{23}$$

In modified gravity frameworks, deviations in $H(z)$ directly propagate into the redshift–luminosity relation, making SNe Ia an especially sensitive probe of cosmic acceleration. For this study, we adopt the Pantheon+SH0ES sample, consisting of 1701 SNe Ia light curves. Following standard practice, we exclude objects with $z < 0.01$ to reduce peculiar-velocity effects, leaving 1588 SNe in the range $0.01 \leq z \leq 2.3$ [5]. The absolute magnitude $M$ is treated as a nuisance parameter.

For the Pantheon+SH0ES dataset, we use the full statistical-plus-systematic covariance matrix $C_{SN}$, which includes calibration uncertainties, light-curve standardization systematics, zero-point offsets, color and stretch variations, and the SH0ES absolute-magnitude calibration. Numerically, assuming uncorrelated uncertainties in the measured data points may lead to overfitting, whereas using the full covariance matrix to account for known correlated systematic errors increases the $\chi^2$ to a more physically plausible value. The residual vector is defined as

$$\Delta\mu_i = m_{B,i} - M - \mu^{\text{th}}(z_i), \tag{24}$$

where $m_{B,i}$ is the observed peak apparent magnitude of the $i$-th supernova and $\mu^{\text{th}}(z_i)$ is the theoretical distance modulus at redshift $z_i$. Following the Pantheon+ methodology, the absolute magnitude $M$ is analytically marginalized. The corresponding chi-squared function is

$$\chi^2_{SN} = \Delta\mu^T C_{SN}^{-1} \Delta\mu. \tag{25}$$

The BAO feature provides yet another independent standard ruler, originating from sound waves in the photon–baryon plasma before recombination. This imprint sets a characteristic scale, the comoving sound horizon at the baryon-drag epoch $z_d$, which is defined as

$$r_s(z_d) = \int_{z_d}^{\infty} \frac{c_s(z)}{H(z)} dz, \tag{26}$$

where $c_s(z)$ is the sound speed in the photon–baryon plasma.

BAO measurements constrain combinations of "radial" Hubble ($D_H$) and "transverse" angular ($D_A$) distances, commonly expressed through the volume-averaged distance ($D_V$):

$$D_V(z) = [z(1+z)^2 D_A^2(z) D_H(z)]^{1/3}, \quad D_H(z) = \frac{c}{H(z)}, \quad D_A(z) = \frac{1}{1+z}\int_0^z dz' \frac{c}{H(z')}. \tag{27}$$

Radial BAO measurements constrain $D_H$, while transverse BAO constraints involve the comoving angular-diameter distance $D_M(z) = (1+z)D_A(z)$.

In this work, we use DESI DR1 BAO measurements, consisting of seven high-precision measurements covering $0.295 \leq z \leq 2.330$ [52]. The measurements are obtained from different tracers, namely the Bright Galaxy Sample (BGS), Luminous Red Galaxies (LRG1–3), Emission Line Galaxies (ELG1–2), Quasars (QSO), and Lyman-α (Lyα) QSOs. For each tracer, we adopt the corresponding published covariance matrices, $C_{BAO}$, to account for correlated uncertainties. The chi-squared function for BAO is constructed as

$$\chi^2_{BAO} = \Delta X^T C_{BAO}^{-1} \Delta X, \tag{28}$$

where $\Delta X$ is the vector of residuals between theoretical predictions and observed values of the relevant BAO observables, namely $D_M(z)/r_s(z_d)$, $D_H(z)/r_s(z_d)$, and $D_V(z)/r_s(z_d)$.

Assuming statistical independence between CC, SNe Ia, and BAO datasets, the total likelihood is given by

$$\chi_{\text{tot}}^2 = \chi_{CC}^2 + \chi_{SN}^2 + \chi_{BAO}^2, \tag{29}$$

with the reduced statistic $\chi_\nu^2 = \chi^2/(N-k)$ (where $N$ is the number of data points and $k$ is the number of model parameters) providing a quantitative measure of the fit quality, and serving as the basis for model comparison.

By combining these probes—direct expansion rates from CC, distance moduli from SNe Ia, and standard-ruler information from BAO—we construct a comprehensive constraint system that minimizes degeneracies and enhances robustness. This integrated framework not only tests each modified gravity scenario against multiple observational fronts but also yields a coherent picture of cosmic expansion, effectively bridging early- and late-time cosmology within a single statistical scheme [49,53].

## Models

5.1. (ACL–G) model

We propose the Affine–Coupled Linear–Geometric (ACL–G) model, a minimal ansatz within the nonminimally coupled $f(T)$ gravity framework. The model is defined by:

$$f_1(T) = \alpha T - 6\xi H_0^2, \qquad f_2(T) = b_0 + b_1 \frac{T}{T_0}. \tag{30}$$

The first function, $f_1(T)$, contains a linear dependence on torsion, with proportionality coefficient $\alpha$, plus a constant term parameterized by $\xi$. The second, $f_2(T)$, is a dimensionless linear function of $T/T_0$, characterized by free parameters $b_0$ and $b_1$, and represents the simplest form of [54]. torsion–matter coupling consistent with dimensional requirements. This construction is intentionally minimal: at early times, where the torsion $T$ is large and matter content $\mathcal{L}_m$ is negligible, the model reduces to standard teleparallel gravity ($f_1 \to \alpha T$, $f_2$ dropped), while at late times the coupling term ($f_2$) can drive accelerated expansion. The free parameters $\{\alpha, \xi, b_0, b_1\}$ can be directly constrained by cosmological data, while theoretical consistency demands a positive effective gravitational coupling and a smooth ΛCDM limit when $(b_1, \xi) \to (0,0)$. This model was inspired from the linear teleparallel dark energy formalism of [36], but the ACL–G formulation extends it by including constant affine terms in both $f_1$ and $f_2$, giving a model that is analytically tractable, observationally testable, and capable of capturing late-time deviations from general relativity in a controlled way. We can determine $\alpha$ by evaluating the functional form given in Eq. (30) at the present time within the first modified Friedmann equation:

$$\alpha = \Omega_{m0}(1 + \lambda b_0 - \lambda b_1) + \xi - 1. \tag{31}$$

By combining Eqs. (8) and (10) with Eq. (30) and the parameter $\alpha$, and subsequently substituting them into Eq. (18), we obtain

$$\frac{\delta \mathcal{T}_f}{\mathcal{T}_f} = \frac{1}{\sqrt{10}} \left[ \left(\frac{\mathcal{T}_0}{\mathcal{T}_f}\right)^4 \xi + (1 + \Omega_{m0}(\lambda(b_1 - b_0) - 1) - \xi) \right] \frac{\pi \sqrt{g_*}}{30q \, M_P \mathcal{T}_f^3}, \tag{32}$$

where by fixing the parameters to $\xi = 0.7$, $b_0 = -0.5$, and $b_1 = 1$, we find that the observationally inferred primordial light element abundances constrain the parameter $\lambda$ to the range $-0.026 \leq \lambda \leq 0.026$, as illustrated in Fig. 1. This bound arises from BBN, which imposes stringent limits on any deviation of the freeze-out temperature from its standard value.

In this analysis, we fix the cosmological parameters to well-established values. Specifically, the present-day density parameters for matter is set to $\Omega_{m0} = 0.315$ [4]. The present CMB temperature is taken as $\mathcal{T}_0 = 2.6 \times 10^{-13}$ GeV [55]. These reference values are used consistently across all models considered in this work.

BBN provides only bounds on $\lambda$, restricting it to an allowed interval rather than fixing its precise value. In order to obtain a tighter constraint, we incorporate the ACL–G framework together with the parameter $\alpha$ into the first Friedmann equation. This leads to the following expression for the dimensionless Hubble parameter, $E(z) \equiv H(z)/H_0$:

$$E^2(z) = \frac{\Omega_{m0}(1+z)^3(1 + \lambda b_0) + \xi}{\Omega_{m0}(1 + \lambda b_0 - \lambda b_1) + \xi + \lambda b_1 \Omega_{m0}(1+z)^3}. \tag{33}$$

Parameter estimation was performed using SNe Ia, BAO, and CC datasets, analyzed both separately and in combination, with the corresponding constraints summarized in Table 1.

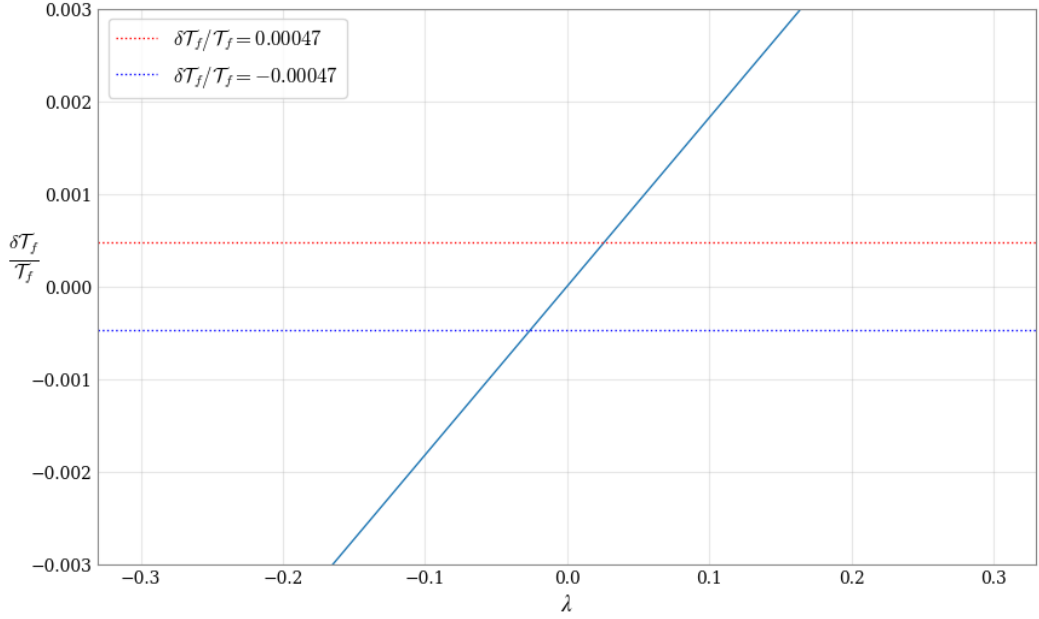

**Figure 1** Fractional change in freeze-out temperature $\delta \mathcal{T}_f/\mathcal{T}_f$ versus $\lambda$. BBN bounds restrict $\lambda$ to the range $[-0.026, 0.026]$.

**Table 1** Best-fit cosmological parameters with $1\sigma$ uncertainties from SNe Ia, BAO, CC, and their combination for the ACL–G model.

| Dataset | $\Omega_{m0}$ | $\xi$ | $\lambda$ | $H_0$ |
|---|---|---|---|---|
| SNeIa | $0.3669^{+0.1520}_{-0.1518}$ | $0.6995^{+0.0346}_{-0.0343}$ | $-0.0014^{+0.0344}_{-0.0330}$ | $67.33^{+0.56}_{-0.24}$ |
| BAO | $0.2822^{+0.0454}_{-0.0392}$ | $0.7019^{+0.0330}_{-0.0349}$ | $-0.0022^{+0.0070}_{-0.0079}$ | $69.51^{+1.21}_{-1.19}$ |
| CC | $0.3356^{+0.0443}_{-0.0425}$ | $0.7016^{+0.0337}_{-0.0350}$ | $0.0144^{+0.0065}_{-0.0069}$ | $68.08^{+0.99}_{-0.72}$ |
| SNe Ia+BAO+CC | $0.3582^{+0.0294}_{-0.0300}$ | $0.7026^{+0.0325}_{-0.0355}$ | $0.0103^{+0.0038}_{-0.0039}$ | $67.55^{+0.60}_{-0.39}$ |

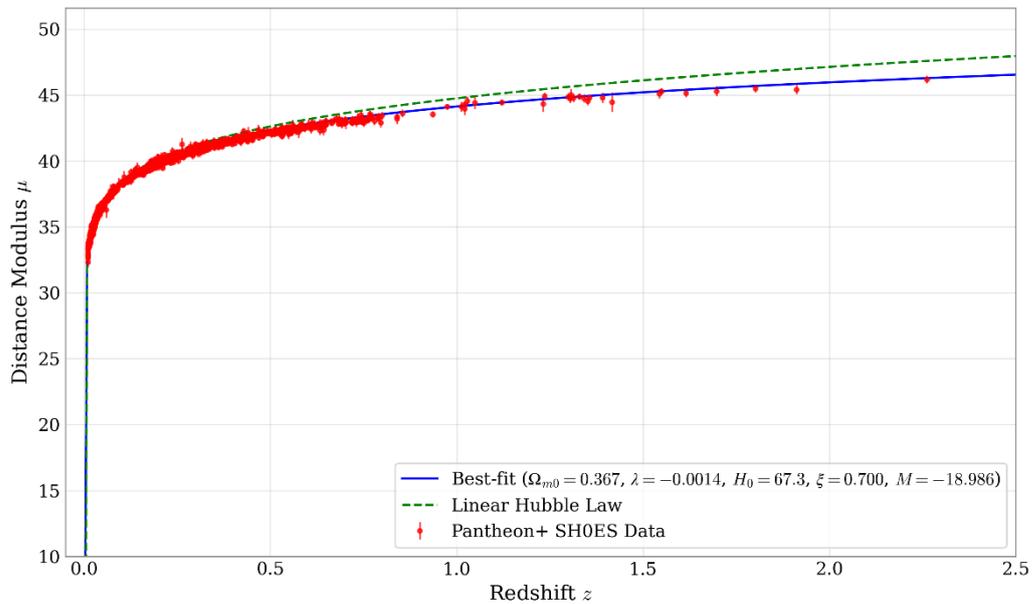

**Figure 2** Distance modulus $\mu$ versus redshift $z$ from Pantheon+SH0ES SNe Ia. The solid line shows the best-fit ACL–G model, whereas the dashed line corresponds to the linear Hubble law.

The goodness of fit is quantified through the reduced chi-squared. For SNe Ia, the model yields $\chi_\nu^2 \simeq 0.2$, a value commonly encountered when using the full Pantheon+SH0ES covariance matrix, where correlated uncertainties broaden the effective error budget. The corresponding constraints are illustrated in Fig. 2, together with the inferred SNe absolute magnitude $M = -18.9858^{+0.6699}_{-0.6974}$. The BAO likelihood produces $\chi_\nu^2 \simeq 1.81$, reflecting both the smaller dataset and the high statistical precision of the DESI measurements (Fig. 3). CC data yield $\chi_\nu^2 \simeq 0.52$, reflecting the typical scatter of current CC measurements and their reliance on differential-age determinations (Fig. 4).

Small differences in the best-fit values of $H_0$ are expected because each observational probe relies on its own calibration method. SNe Ia are sensitive to the absolute magnitude $M$, BAO measurements use the sound horizon, and CC estimate $H(z)$ directly from galaxy ages, which introduces more noise.

These probe-specific differences are naturally reconciled in the joint SNe Ia + BAO + CC analysis, which yields a reduced chi-squared of $\chi_\nu^2 \simeq 0.3$ and produces a consistent combined estimate of the cosmological parameters, including $M = -19.0148^{+0.6969}_{-0.6756}$. This result demonstrates that the model offers a stable and coherent description of the cosmic expansion history across all three observational probes (Fig. 5).

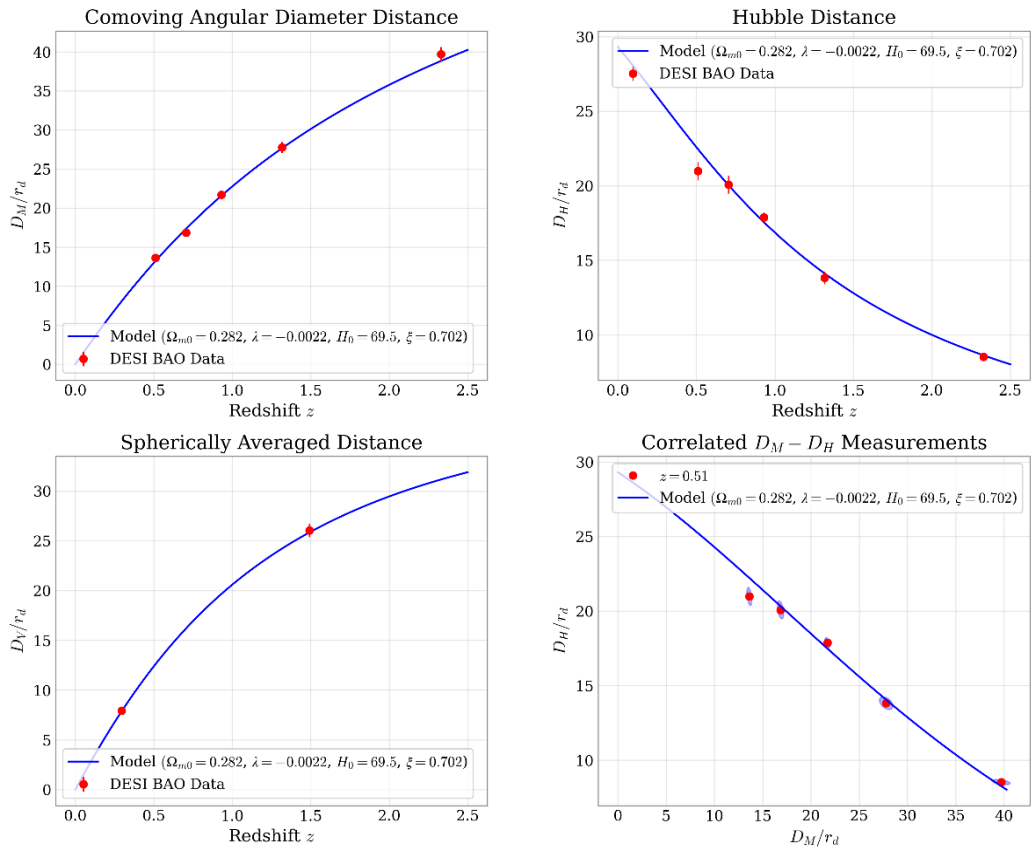

**Figure 3** DESI BAO measurements compared with the best-fit prediction of the ACL–G model. Top-left: comoving angular diameter distance $D_M/r_d$ as a function of redshift. Top-right: Hubble distance $D_H/r_d$. Bottom-left: spherically averaged BAO distance $D_V/r_d$. Bottom-right: joint constraints in the $D_M/r_d$, $D_H/r_d$ plane. Red points with error bars show DESI BAO data, and the blue curves represent the ACL–G model prediction. The model provides a consistent fit across all BAO observables.

Examining the individual datasets in more detail, $\Omega_{m0}$ exhibits mild tension: SNe Ia tend to prefer slightly higher values, while BAO favors lower ones. The joint fit, however, converges to an intermediate value that remains fully consistent within the $1\sigma$ uncertainties, highlighting the constraining power of combining multiple probes. The parameter $\xi$ is remarkably stable at $\sim 0.70$ across all datasets, indicating a weak dependence on probe-specific calibrations. The most interesting behavior is observed in $\lambda$: SNe Ia and BAO mildly favor negative values, whereas CC strongly prefers a positive value. Consequently, the joint likelihood is driven toward a clearly positive $\lambda$, reflecting the sensitivity of CC measurements to the slope of $H(z)$ at intermediate redshifts. The inferred values of $H_0$ follow the expected trends for each probe, with the combined fit providing a balanced and robust estimate that is fully consistent across the datasets.

Overall, the joint analysis confirms that the model is capable of simultaneously accommodating the subtle differences inherent in each probe while delivering a coherent picture of the cosmic expansion history.

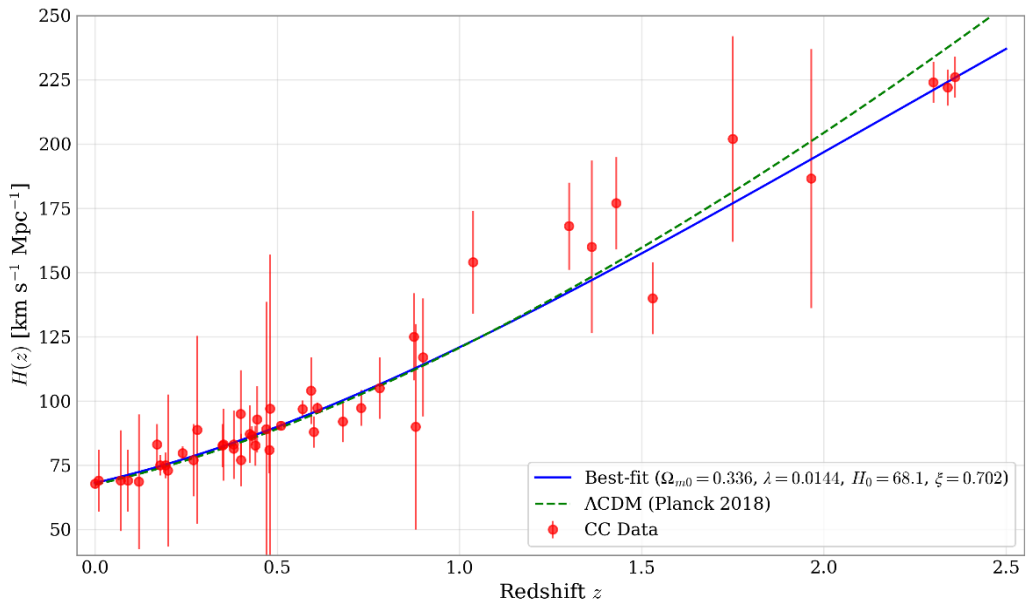

**Figure 4** CC measurements of H(z) compared with the best-fit prediction of the ACL–G model. The red points with error bars show the CC data, the blue solid curve denotes the ACL–G best-fit evolution of the Hubble parameter, and the green dashed curve shows the ΛCDM prediction based on Planck 2018. The ACL–G model closely follows the observational trend across the full redshift range.

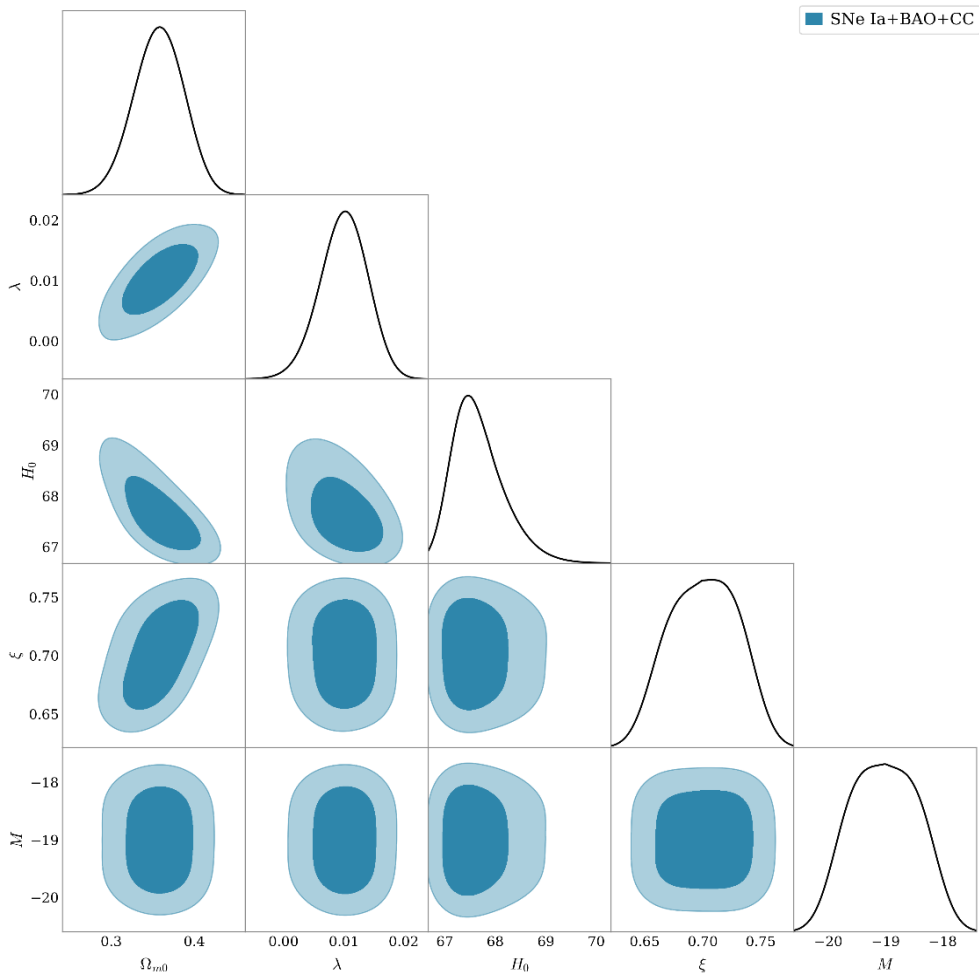

**Figure 5** Joint posterior distributions for $\Omega_{m0}$, $\lambda$, $H_0$, and $\xi$ for the ACL–G model, derived from the combined SNe Ia+BAO+CC analysis, displaying the 68% and 95% confidence contours.

## 5.2. Power–Law

We introduce the Power–Law model, defined by

$$f_1 = \beta_1(-T)^n, \qquad f_2 = \beta_2(-T)^r, \tag{34}$$

where $\beta_1$ and $\beta_2$ are free coupling coefficients and $n$ and $r$ are real indices controlling the nonlinearity of the gravitational and matter–torsion sectors, respectively.

This form generalizes the successful power–law modifications widely explored in $f(R)$ gravity and $f(T)$ cosmology, while extending the nonminimal torsion–matter coupling approach of [54] to allow independent scaling in $f_1$ and $f_2$. The parameter $\beta_1$ controls the amplitude of the pure torsion modification, with $n = 1, \beta_1 = 1$ recovering the TEGR, while $\beta_2$ and $r$ dictate the strength and scaling of the matter–torsion coupling, with $r > 0$ ($r < 0$) enhancing the coupling in the early (late) universe. Viable parameter regions must satisfy a positive effective gravitational coupling $G_{\text{eff}} > 0$ over the entire redshift range, reduce smoothly to ΛCDM in the limit $\beta_2 \to 0, n \to 1, \beta_1 \to 1$, and remain consistent with BBN bounds, which typically constrain $|\beta_2|$ for large positive $r$. Observationally, $\beta_1$ and $n$ exhibit partial degeneracy with $\Omega_{m0}$, while $\beta_2$ and $r$ can mimic dynamical dark–energy behavior, introducing degeneracy with $H_0$ and the effective equation–of–state parameter. One can obtain $\beta_1$ by using Power–Law form (34) in Eq. (6) at the present time:

$$\beta_1 = \frac{\Omega_{m0}[1 + \lambda\beta_2(1 + 2r)(6H_0^2)^r] - 1}{(1 + 2n)(6H_0^2)^{n-1}}, \tag{35}$$

so plugging Eq. (34) and Eq. (10) into Eq. (18), we obtain:

$$\frac{\delta \mathcal{T}_f}{\mathcal{T}_f} = \sqrt{\frac{\pi^2 g_*}{90} \frac{(\Omega_{m0}[1 + \beta_2[\lambda(1 + 2r)]H_0^{2r}] - 1)}{10q\, M_P \mathcal{T}_f^3}} \left(\frac{\mathcal{T}_f}{\mathcal{T}_0}\right)^{4n-4}. \tag{36}$$

Compatibility with BBN constraints can be maintained provided that $n \leq 0.98$, as illustrated in Figure 6, where we fix the parameters $r = -0.01$ and $\beta_2 = 1$. The BBN limits on the relative shift in the freeze-out temperature, $\delta\mathcal{T}_f/\mathcal{T}_f$, permit a rather broad interval for $n$. Consequently, BBN alone does not determine the precise value of $n$.

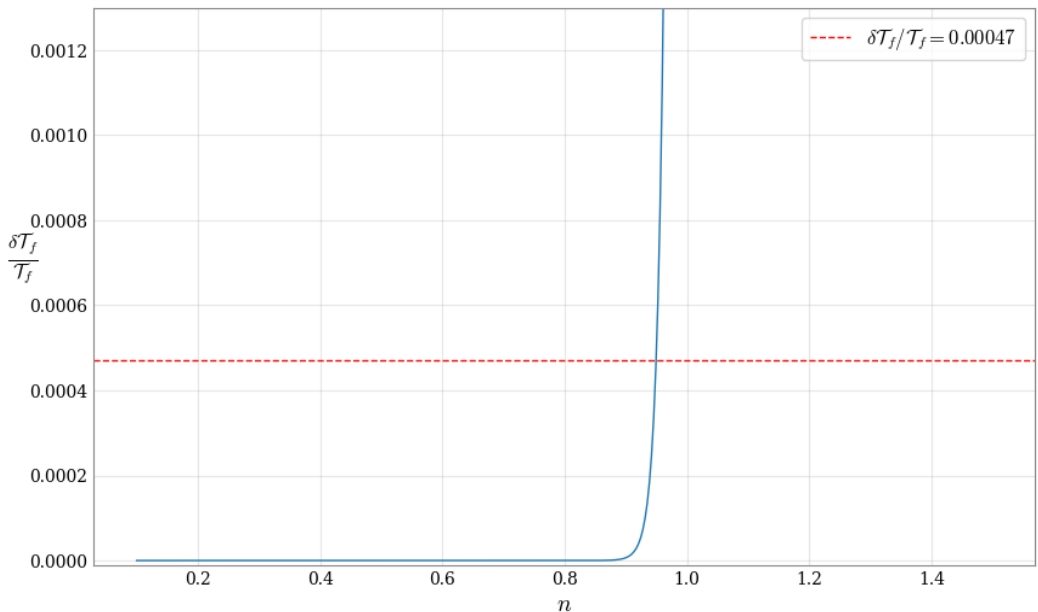

**Figure 6** The variation of the relative freeze-out temperature shift, $\delta\mathcal{T}_f/\mathcal{T}_f$ versus $n$. The BBN bound intersects at $n \approx 0.98$, thereby enforcing the condition $n \leq 0.98$ as the observationally acceptable region.

To address this issue, we substitute Eq. (34) together with Eq. (35) into Eq. (6), which yields the following modified Friedmann equation:

$$E^2(z) = \Omega_{m0}(1 + z)^3[1 + CE^{2r}] + [1 - \Omega_{m0}(1 + C)]\, E^{2n}, \tag{37}$$

where $C \equiv \lambda(1+2r)\beta_2\, 6^r H_0^{2r}$. In order to test the consistency of the Power–Law model, parameter estimation was performed using independent observational datasets. The resulting best-fit values are summarized in Table 2, while the reduced chi-squared values $\chi_\nu^2$ are discussed separately.

**Table 2** Best-fit cosmological parameters with $1\sigma$ uncertainties from SNe Ia, BAO, CC, and their combination for the Power–Law model.

| Dataset | $\Omega_{m0}$ | $n$ | $\lambda$ | $H_0$ |
|---|---|---|---|---|
| SNe Ia | $0.4637^{+0.0954}_{-0.1402}$ | $0.1073^{+0.2850}_{-0.3837}$ | $-0.2969^{+0.1379}_{-0.1808}$ | $64.38^{+0.62}_{-0.29}$ |
| BAO | $0.4328^{+0.0984}_{-0.0814}$ | $0.0031^{+0.2468}_{-0.2648}$ | $-0.3758^{+0.1817}_{-0.1493}$ | $69.35^{+1.54}_{-1.62}$ |
| CC | $0.3789^{+0.0947}_{-0.0740}$ | $0.3149^{+0.1241}_{-0.1755}$ | $-0.3827^{+0.1887}_{-0.1558}$ | $68.07^{+1.10}_{-1.05}$ |
| SNe Ia+BAO+CC | $0.4084^{+0.0990}_{-0.0767}$ | $0.3924^{+0.0737}_{-0.1125}$ | $-0.3782^{+0.1821}_{-0.1547}$ | $66.79^{+0.77}_{-0.73}$ |

The SNe Ia dataset yields an excellent fit with $\chi_\nu^2 = 0.19$, consistent with expectations when the full Pantheon+SH0ES covariance structure is included. The corresponding constraints are shown in Fig. 7, together with the inferred absolute magnitude $M = -19.0066^{+0.6996}_{-0.6747}$. The BAO likelihood gives $\chi_\nu^2 = 1.83$, reflecting the tighter statistical weight of DESI measurements and the smaller sample size (Fig. 8). CC observations yield $\chi_\nu^2 = 0.53$, in line with their heterogeneous uncertainties and the nearly calibration-independent nature of differential-age measurements (Fig. 9).

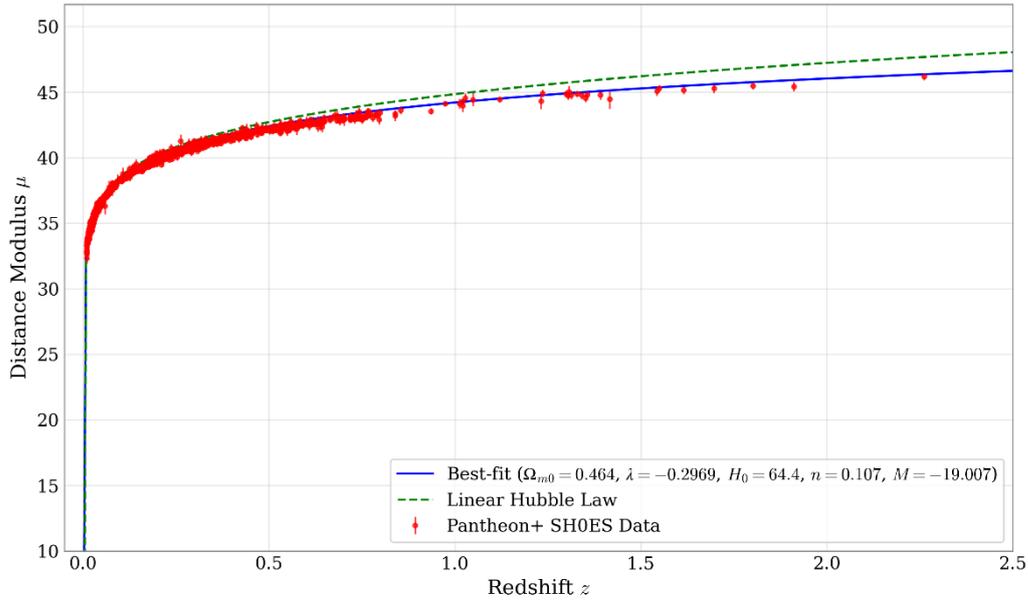

**Figure 7** Distance modulus $\mu$ as a function of redshift $z$ using Pantheon+SH0ES SNe Ia. The blue solid curve corresponds to the best-fit Power–Law model, compared with green dashed curve corresponding to the linear Hubble law.

As in the linear model, slight differences in the individually inferred values of $H_0$ are expected, since each dataset probes the late-time expansion with different calibrations and parameter degeneracies. In the joint SNe Ia + BAO + CC fit, these effects are naturally balanced, yielding a consistent combined constraint with $\chi_\nu^2 = 0.22$ and $M = -19.0215^{+0.7096}_{-0.6691}$ (Fig. 10), confirming the coherence of the power-law model across all probes.

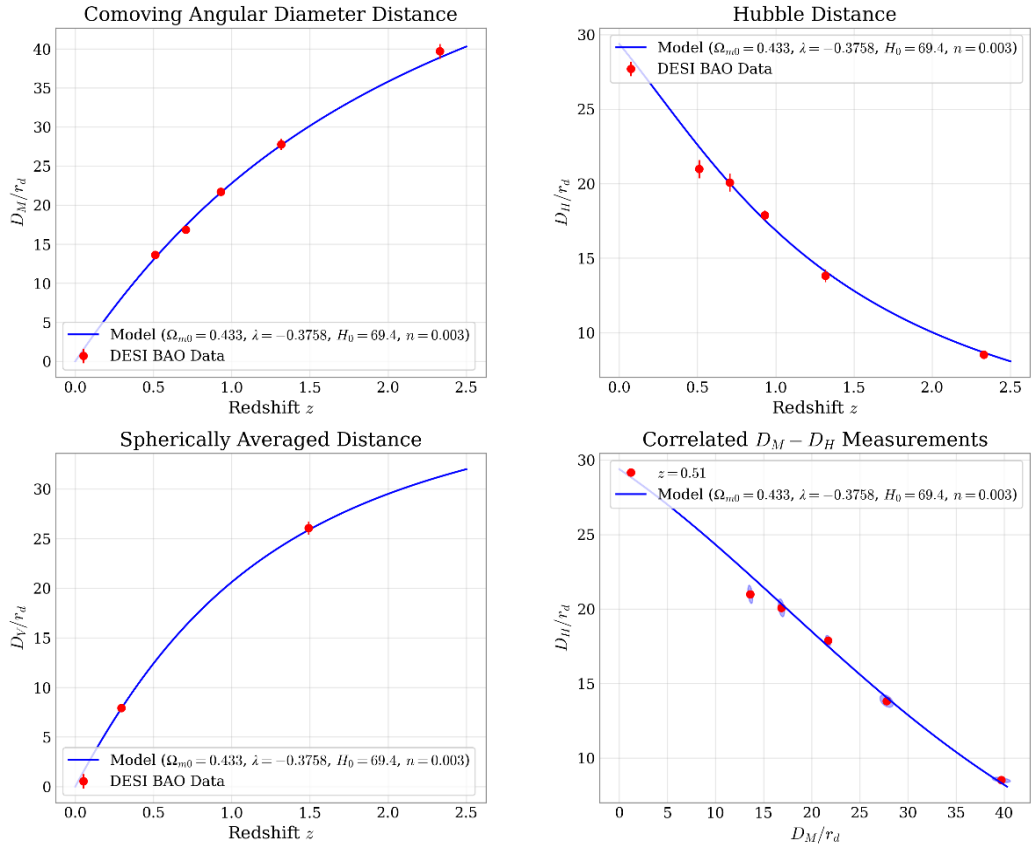

**Figure 8** DESI BAO distance measurements (red markers) compared with the Power–Law model (blue) for comoving angular diameter distance, Hubble distance, spherically aver-aged distance, and correlated $D_M - D_H$ relations.

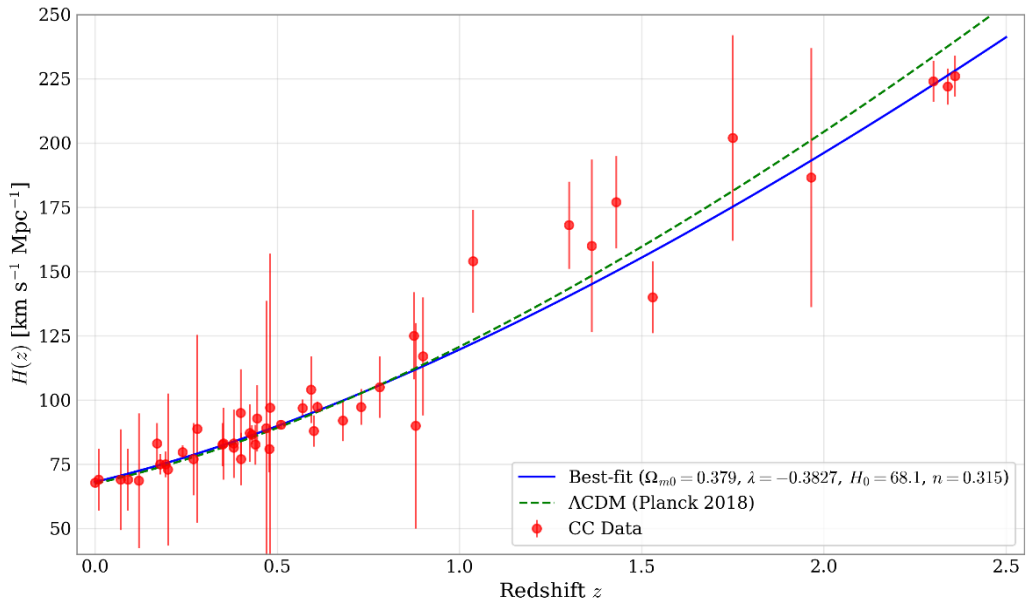

**Figure 9** CC measurements from cosmic chronometers (red points) together with the best-fit Power–Law prediction (blue line) and the ΛCDM curve from Planck 2018 (green dashed).

The combined constraints drive $\lambda$ and $n$ close to zero, pointing to mild yet non-vanishing deviations from minimal torsion–matter coupling. The model thus remains theoretically stable while offering an excellent fit to observations. Compatibility with BBN bounds further restricts the parameter space, while late-time probes from SNe, BAO and CC provide mutually consistent $H_0$ estimates. Altogether, the analysis indicates that departures from minimal coupling are small, consolidating the model's viability as a description of cosmic expansion consistent with both early- and late-time data.

Across the individual datasets, $\Omega_{m0}$ shows a clear trend: SNe Ia and BAO favour higher values, CC prefers a slightly lower one, and the joint fit settles at an intermediate value consistent within uncertainties. The parameter $n$ displays noticeable dataset dependence—BAO prefers values near zero, CC favours a positive deviation, and SNe Ia allow a broad range—so the combined result reflects the CC-driven tendency toward $n > 0$. The parameter $\lambda$ is consistently negative across all datasets, indicating a strong and uniform preference for $\lambda < 0$, which carries directly into the joint constraints. Regarding the expansion rate, SNe Ia favour a distinctly lower $H_0$, BAO shift the preferred value upward, and CC occupy an intermediate position, while the joint fit naturally converges to a balanced value consistent with all probes.

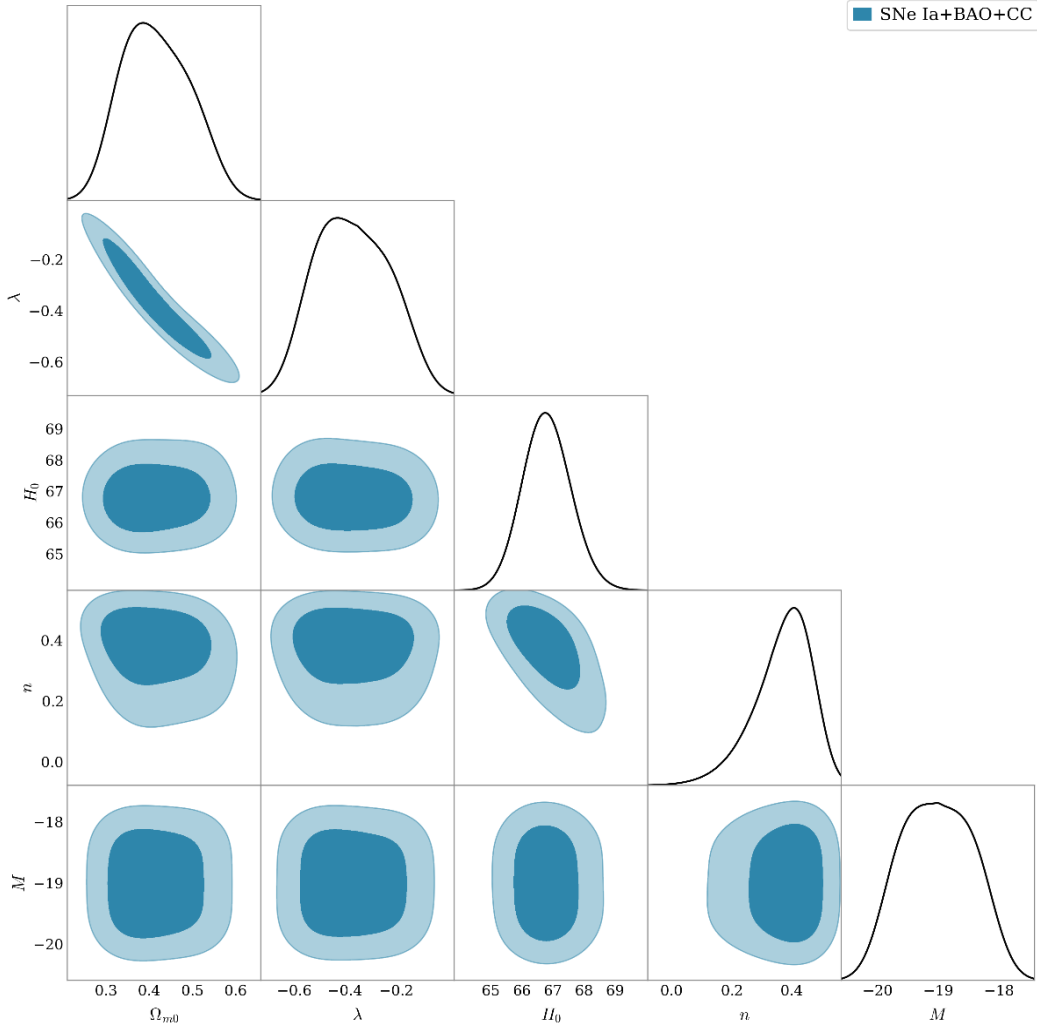

**Figure 10** Marginalized posterior distributions of $\Omega_{m0}$, $\lambda$, $H_0$, and $n$ for the Power–Law scenario using the combined SNe Ia+BAO+CC dataset (68% and 95% C.L. regions).

## 6. Conclusion

We have developed and implemented a self-consistent, multi-epoch testing framework for $f(T)$ gravity with a nonminimal torsion–matter coupling, combining early-universe constraints from Big Bang Nucleosynthesis with late-time expansion probes including type Ia supernovae, baryon acoustic oscillations, and cosmic chronometers. The core of this approach lies in confronting the torsion–matter coupling not in isolation, but with a consistent cross-epoch analysis that probes the theory's predictions across the entire thermal and expansion history of the Universe. Applying this framework to two representative models—the linear ACL–G model and a power-law model—we have shown that BBN already imposes meaningful bounds on the coupling strength, and that late-time probes further refine (and in some cases shift) these constraints in a statistically coherent way.

For the ACL–G model, the combined SNe Ia+BAO+CC fit yields a clearly positive best-fit value for the coupling parameter $\lambda$, even though SNe Ia and BAO data separately favour negative values. This sign reversal emerges only when all datasets are combined, underscoring a key advantage of our multi-probe methodology: it can uncover parameter trends that individual datasets cannot robustly determine. The stability of the parameter $\xi \sim 0.70$ across all late-time probes, along with a consistent estimate of $H_0$, indicates that the model accommodates late-time observations without internal inconsistencies. Notably, the joint fit produces a low reduced $\chi^2$ and returns parameter values fully compatible with the BBN prior $|\lambda| \leq 0.026$, confirming the model's viability from BBN to the present day.

The power-law model is likewise constrained effectively by this multi-epoch approach. BBN restricts viable models to $n \leq 0.98$, and the late-time analysis drives both $n$ and $\lambda$ toward zero, indicating only mild deviations from minimal torsion–matter coupling. While individual datasets display strong parameter disparities—especially in $n$—the combined constraints converge on a consistent region that satisfies both early- and late-time requirements. The excellent goodness of fit achieved in the joint analysis, together with compatibility with BBN limits, reinforces the robustness of the power-law model under multi-epoch scrutiny.

In summary, the main result of this study is the demonstration that nonminimally coupled $f(T)$ models can be constrained simultaneously by early-universe nuclear physics and late-time precision cosmology within a single, coherent statistical framework. This constitutes a significant methodological advance over earlier analyses that treated these epochs separately. Our results indicate that viable departures from minimal coupling are small but non-zero, and that combined information from SNe Ia, BAO, CC, and BBN yield concordant cosmological parameters without tension in $H_0$, $\Omega_{m0}$, or the coupling sector. Looking ahead, the framework developed here provides a natural basis for future extensions, including the incorporation of full-shape large-scale-structure data, gravitational-wave standard sirens, or early-universe probes beyond BBN such as CMB anisotropies. Such developments would enable more stringent, multi-probe tests of nonminimal torsion–matter interactions and clarify whether the small deviations identified here reflect genuine new physics or simply the upper bounds of viable modifications within teleparallel gravity.

**Funding:** This research did not receive any external funding.

**Data Availability Statement:** The data supporting the findings of this study are available within the article.

**Conflicts of Interest:** The authors declare no conflict of interest.